# High current DyBCO- ROEBEL Assembled Coated Conductor (RACC)

**W Goldacker, R Nast, G Kotzyba, S I Schlachter, A Frank, B Ringsdorf, C Schmidt, P Komarek**

Forschungszentrum Karlsruhe, Institut for Technical Physics, P.O. Box 3640, D-76021 Karlsruhe

E-mail: wilfried.goldacker@itp.fzk.de

**Abstract**. Low AC loss high transport current HTS cables (>1 kA) are required for application in transformers, generators and are considered for future generations of fusion reactors coils. 2G coated conductors are suitable candidates for high field application at quite high operation temperatures of 50-77 K, which is crucial precondition for economical cooling costs. As a feasibility study we present the first ROEBEL bar cable of approx. 35 cm length made from industrial DyBCO coated conductor (THEVA GmbH, Germany). Meander shaped ROEBEL strands of 4 mm width with a twist pitch of 180 mm were cut from 10 mm wide CC tapes using a specially designed tool. The strands carried in average 157 Amps/cm-width DC and were assembled to a subcable with 5 strands and a final cable with 16 strands. The 5 strand cable was tested and carried a transport current of > 300 Amps DC at 77 K, equivalent to the sum of the individual strand transport critical currents. The 16 strand cable carried 500 A limited through heating effects and non sufficient stabilisation and current sharing. A pulse current load indicated a current carrying potential of > 1 kA for the 16 strand cable.

## 1. Introduction
Coated conductors (CC) are handled as the second generation (2G) of HTS superconductors with the prospects to be 3-5 times cheaper than BSCCO tapes and having the advantage of an application regime at significantly higher fields of a few TESLA at 77 K and in the high field regime >10 T at temperatures around 50 K. This favors CC for the application in transformers, motors, generators and high field coils. Future HTS coils for Fusion reactors will dramatically reduce the cooling costs compared to the actually applied LTS coils operated at 4 K. CC are developed with a tape width in the range of 4-12 mm and a thickness of approximately 0.1 mm made by a large variety of preparation methods and more or less complicated architectures of the layers. The actual global activities focus on getting high current densities over long lengths in a reliable way, to improve the preparation speed and to simplify the architecture to fewer layers. Actually up to 212 m CC length with 245 A transport current and preparation speeds in the range of 3-10 m/h per layer are reached [1,2]. The superconductor is generally applied as an unstructured monolayer which is very unfavorable for low AC losses due to the high aspect ratio of the cross section. Striation experiments structuring the YBCO-layer showed the expected reduction of the AC losses [3]. Much more important for low AC losses is the realization of a twisted superconducting percolation path, however no practical solutions are known so far to realize this in a CC. Since the application of CC in devices requires often higher transport currents (>1 kA) than the actually reached 200-480 A/cm-width in single tapes, low AC loss cable structures are needed. Beside Rutherford cables, ROEBEL bars are the mostly suited cable arrangements and are already successfully applied as NbTi ROEBEL cable in the LCT-coil [4], a fusion reactor model coil, and as HTS BSCCO ROEBEL bar for a transformer [5]. In this paper the first preparation of a fully transposed ROEBEL bar from CC is presented. Goal of the work is showing the feasibility of the technique in principle.

## 2. ROEBEL bar concept

The first and so far only application of ROEBEL bars with HTS conductors was realized by SIEMENS with a 13 strand (insulated) BSCCO(2223) cable for an application in transformers, motors and generators [5]. This concept was successful in performing a high amperage conductor (400 A, 77K, self field) with a transposition length of 3 m. Self field effects reduced the current (DC) of the cable to about 60-70% compared to the sum of the strand currents. The very poor in plane bending capability of the BSCCO tape hinders to realize a twist pitch of the strands in the decimeter regime.

Since coated conductor tapes possess equivalent limitations for in plane bending, the new concept of pre-shaping a ROEBEL strand structure from single CC tapes followed by an assembling process (RACC=ROEBEL Assembled CC) was developed (see Fig.1-2). Keypoint of the idea is that the quite thin CC can be cut mechanically into a ROEBEL strand shape, so far already used to split a coated conductor into smaller straight stripes. For the presented feasibility study the twist pitch of the ROEBEL strands was chosen as 180 mm, the step over angle to 30°, the strand width to 4 mm. A cable length of 2 twist pitch lengths was prepared to prove the relevance of the assembling procedure for longer lengths.

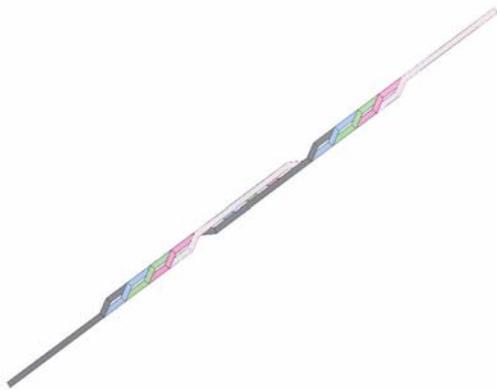 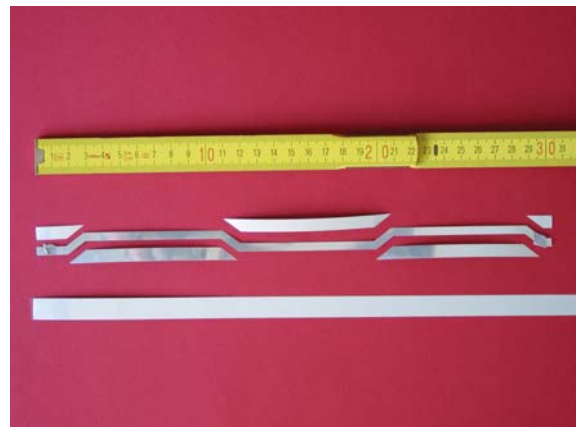

**Figure 1.** Sketch of a 5 strand ROEBEL bar    **Figure 2.** CC tape (bottom) cut to ROEBEL strand

## 3. Experimental

As basic materials THEVA DyBCO-tapes were used [6]. The architecture consisted of 0.09 mm Hastelloy substrate, 3 microns ISD MgO buffer, 1.8 microns DyBCO and 400 or 200 nm Ag cap layer. End-to-end (3.9 and 10 m) transport critical currents (77K, self field) were > 200 A. The transport current of a short sample (5 cm) of the CC was investigated, with and without Cu stabilizer (50 micron thickness) soldered on the cap layer testing the functionality of the thermal stabilisation.

For the RACC-cable 40 cm CC tape pieces were measured and then cut into the 38 cm long ROEBEL strand structure (see fig.2), by means of a specially designed mechanical tool. Transport currents of raw tape and strand were measured. Current contacts were applied with pressed Indium foil, voltage taps worked with spring tips.

Assembling of the cable was made around an insulated central Constantan tape as mechanical support, successively winding the strands. In the first stage of a subcable 5 CC strands with 400 nm Ag cap layer and 1 Cu strand (0.25 mm thickness) were applied (see Fig. 4).

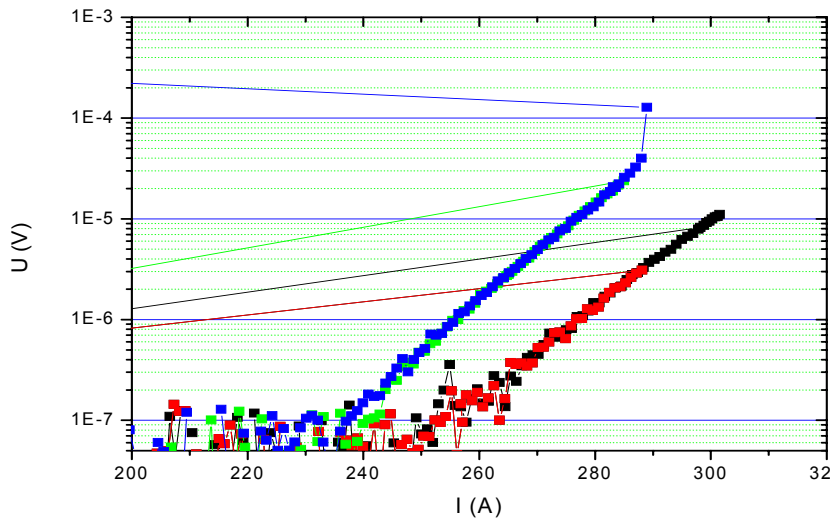

**Figure 3.** Transport critical current measurement of a 5 cm DyBCO-CC tape with 400nm Ag cap layer (blue line) at 77K and in self field and with additional 50 microns Cu layer (red and black line) soldered on the Ag cap layer

In the second step 11 additional strands were added, 10 with 200 nm Ag cap. Neither an interstrand connection nor an insulation between the strands was applied. Current contacts to the cable were performed with Indium foil stapled and pressed between the strand ends (5 mm contact length). Voltage tabs were located on one selected high current strand with spacing of one twist pitch length.

## 4. Results

The transport critical current measurement of a typical short sample shown in Figure 3 with and without additionally soldered Cu stabilisation shows significantly higher values of 255 A and 277 A compared to the whole batch lengths, indicating some inhomogeneity of the tapes. Only stabilized by the Ag cap, the transition is thermally driven above 30 µV/cm. Adding copper an improved transport current and current sharing between CC and copper sheath is observed.

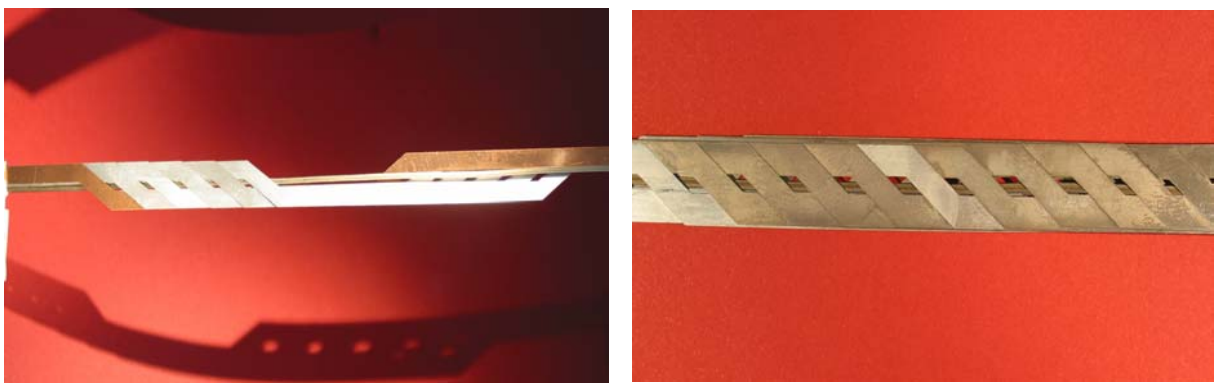

Figure 4. View to a section of the 5 CC + copper strand subcable (left side) and the 16 strand RACC cable with parallel underlying copper stabilisation (5 mm$^2$ cross section, not visible in figure).

The statistics of the transport current investigations of tape sections, ROEBEL strands and cables is given in the table. The 16 tape sections had average transport currents of 232 A, the ROEBEL strands 63 A (equiv. 157 A/cm-width) which is a quote of 68% remaining currents in the strands. This is an excellent value since the known current inhomogeneity, anisotropic currents in the tape plane and a non optimized strand geometry contribute to the reduced strand performance. Optically no damage was observed from cutting.

In the 5 strand cable with 400 nm Ag cap a 300 A DC transport current was measured (=current supply limit) which is slightly higher than the calculated current of 294 A caused by the copper strand in parallel. The current redistribution and sharing between the different strands and the Cu worked quite well. For the 16 strand cable with 10 strands having a thinner cap layer a transport current of 500 A was measured, about half the value (1008 A) being calculated (see table). The U(I) transition shows an instable behaviour indicating severe problems of current redistribution.

In a subsequent experiment a current supply function failure caused a sudden overload of the cable with the current supply maximum of 1500 A for a few seconds. Surprisingly the cable kept superconducting the whole sweep up to 1500 A. When the down sweep crossed 1100 A the cable burned through from the meanwhile overheating. This unplanned "pulse current experiment" with much higher current load gave the valuable information that thermal instabilities and heating effects are the main limitations for the cable performance. An optimised thermal stabilisation and cable design obviously would allow transport currents > 1000 A which is consistent with the short sample performance. To confirm this last result and interpretation, detailed investigations are under way.

| Sample | $I_c$ (77K, s.f.) measured (calculated) | Sample width | Comments |
|---|---|---|---|
| CC* /Cu ( l=5 cm ) | 277 A | 10 mm | Cu 50μm soldered |
| CC ( l=40 cm ) | 232 A | 10 mm | 16 tapes average |
| R-strand (l=38 cm) | 63 A = 157 A/cm-width | 4 mm | 16 strand average |
| RACC 5 strands | >300 A ( 294 A ) | 5 x 4 mm | Cu 1 mm$^2$ parallel |
| RACC 16 strands | 500 A** ( 1008 A) | 16 x 4 mm | Cu 5 mm$^2$ parallel |

**Table:** DC transport critical current measurements of CC tapes, strands and RACC-cable. The applied criterion was 0.1 μV/cm (* 1 μV/cm). **This value increased to 1500 A for a pulse current load of a few seconds duration (see text).

## 5. Conclusions and Outlook

The first ROEBEL assembled coated conductor (RACC) cable with a length of 35 cm and twist pitch of 18 cm was presented, the feasibility of the technique was demonstrated and the capability to realize transport currents of >1kA at 77 K was shown. The technique to pre-shape ROEBEL bar strands followed by cable assembling (RACC-technique) tends out to be compatible with the tolerable bending load of the CC material. A sufficient thermal stabilization of the strands is crucial and is a central point of actual improvements. A moderate conductive coupling of the strands is considered for the next cable and AC loss investigations. A future improved current homogeneity in the coated conductor (in work) will strongly diminish current redistribution problems in the RACC cable.

The new RACC cable design for coated conductors opens the possibility for CC low AC loss cables with a current carrying capacity of several kA, which is required for the application in devices like transformers, motors/generators and magnets and in particular for future use in the next generation of superconducting magnets in accelerator facilities and fusion reactors.

**Acknowledgement:** This work was in parts supported by EFDA

**References**
[1] Y.Yamada, T.Watanabe, T.Muroga, S.Miyata, H.Iwai, A.Ibi, K.Takahashi, H.Kobayashi, M.Konishi, Y.Shiohara, T.Katoh, T.Hirayama, ICMC2005, Keystone (USA) M1-G-01, to be published in Adv. Cryog. Eng.
[2] Y.Qiao. Y.Chen, X.Xiong, Y.Li, P.Hou, Y.Y.Xie, J.L.Reeves, T.M.Salagaj, A.R.Knoll, K.P.Lenseth, V. Selvamanickam, ICMC2005 Keeystone (USA) M1-G-03, to be published in Adv. Cryog. Eng.
[3] N.Amemiya, S.Kasai, K.Yoda, Z.Jiang, G.A.Levin, P.N.Barnes, Ch.Oberly Superc. Sci.Technol. 17 (2004) p.1464
[4] The IEA Large Coil Task, eds. D.S.Beard, W.Klose, S.Shimamoto, G.Vecsey, Fusion Engineering and design 7 (1988) p.23-53
[5] V.Hussenether, M.Oomen, M.Leghissa, H.-W-Neumueller, Physica C 401 (2004) p.135
[6] W.Prusseit, G. Sigl, R. Nemetschek, C. Hoffmann, J. Handke, A. Luemkemann, H. Kinder IEEE Transactions on Applied Superconductivity, Vol. 15, No. 2, June2005, p.2608-2610